\documentclass[12pt]{article}
\usepackage{lineno}
\RequirePackage{amsmath}
\RequirePackage{physics}
\RequirePackage{graphicx,verbatim}
\RequirePackage{placeins}
\RequirePackage[utf8]{inputenc} 
\RequirePackage[numbers,sort&compress]{natbib}
\newcommand{\dyturbo}{\texttt{DYTurbo}}

\newcommand{\hturbo}{\texttt{HTurbo}}
\newcommand{\hres}{\texttt{HRes}}
\newcommand{\hqt}{\texttt{HqT}}
\newcommand{\hnnlo}{\texttt{HNNLO}}

\newcommand{\vegas}{Vegas}

\newcommand*{\pT}{\ensuremath{p_{\mathrm{T}}}}

\newcommand*{\qT}{\ensuremath{q_{\mathrm{T}}}}

\newcommand*{\qt}{\ensuremath{q_{\mathrm{T}}}}
 
\newcommand\as{\ensuremath{\alpha_{\mathrm{S}}}} 
\usepackage{subfigure}
\usepackage{booktabs}
\usepackage{microtype}
\begin{document}

\begin{titlepage}
\begin{flushright}
FTUV-22-0218.4393\\
IFIC/22-07
\end{flushright}

\renewcommand{\thefootnote}{\fnsymbol{footnote}}
\vspace*{1cm}

\begin{center}
{\Large \bf 
  Higgs boson production at the LHC:  \\  [0.2cm]
  fast and precise predictions in QCD at higher orders
}
\end{center}

\par \vspace{2mm}
\begin{center}
  {\bf Stefano Camarda${}^{(a)}$},  {\bf Leandro Cieri${}^{(b)}$},\\
 {\bf Giancarlo Ferrera${}^{(c)}$} and {\bf Jes\'us Urtasun-Elizari}${}^{(c)}$\\

\vspace{5mm}

${}^{(a)}$ 
CERN, CH-1211 Geneva, Switzerland\\\vspace{1mm}

${}^{(b)}$ 
Instituto de F\'isica Corpuscular, Universitat de Val\`encia - CSIC\\
Parc Cient\'ific, E-46980 Paterna, Valencia, Spain\\\vspace{1mm}

${}^{(c)}$ 
Dipartimento di Fisica, Universit\`a di Milano and\\ INFN, Sezione di Milano,
I-20133 Milan, Italy\\\vspace{1mm}

\end{center}

\vspace{1.5cm}

%
\par \vspace{2mm}
\begin{center} {\large \bf Abstract} \end{center}
\begin{quote}
  \pretolerance 10000
 
We present a new numerical program, {\ttfamily HTurbo}, which provides fast and numerically precise predictions
for Higgs boson production cross sections.
The present version of the code implements the perturbative QCD expansion up to the next-to-next-to-leading
order also combined  with the resummation of
the large logarithmic corrections at small transverse momenta up to next-to-next-to-leading logarithmic
accuracy
and it includes  the Higgs boson production through gluon fusion and decay in two photons with
the full dependence on the final-state kinematics.
Arbitrary kinematical cuts can be applied to the final states
in order to obtain fiducial cross sections and associated kinematical
distributions.
We present a benchmark comparison with
the predictions obtained with the
numerical programs 
\hres{} and \hnnlo{}  programs
for which \hturbo{}  represents an improved reimplementation.

\end{quote}

\vspace*{\fill}
\vspace*{2.5cm}

\begin{flushleft}
February 2022
\end{flushleft}
\end{titlepage}

\setcounter{footnote}{1}
\renewcommand{\thefootnote}{\fnsymbol{footnote}}

After
the discovery of the Higgs boson~\cite{ATLAS:2012yve,CMS:2012qbp},
a foremost goal of the physics program at the Large Hadron Collider
(LHC) has become
the direct investigation of the electroweak symmetry breaking mechanism.
In particular, precision studies are a key tool
for searches of possible deviations from from Standard Model (SM) predictions
in the Higgs sector.

In this paper we consider the production 
of the SM Higgs boson through gluon fusion and its decay
to $\gamma\gamma$.
The gluon fusion subprocess $gg \to H$, through a heavy-quark loop,
is the main production mechanism of the 
SM Higgs boson at hadron colliders and
its dynamics is driven by strong interactions.
It is thus essential 
to study the effect of QCD radiative corrections at higher orders
and to provide accurate theoretical predictions and
for Higgs boson cross sections and
associated distributions.

The QCD radiative corrections to the total cross section have been computed
up to
next-to-next-to-next-to-leading order (N$^3$LO)
in Refs.~\cite{Anastasiou:2015vya,Dulat:2018bfe,Cieri:2018oms,Billis:2021ecs,Chen:2021isd}. 
The NNLO 
parton-level calculations of $H$+jet production
have been computed in
Refs.~\cite{Chen:2014gva,Boughezal:2015dra,Boughezal:2015aha}.
The combination of the resummation formalism of logarithmically enhanced
contribution at small \qt{} in QCD
with fixed-order perturbative results at different levels of
theoretical accuracy, have been obtained in~\cite{Gao:2005iu,Cao:2007du,deFlorian:2012mx,Bizon:2018foh,Rabemananjara:2020rvw}
(see also references therein).

Theoretical predictions  depend on
several different parameters and on various inputs
such as parton density functions (PDFs),
renormalization and factorization scales and SM parameters.
In order to obtain precise theoretical predictions with a reliable estimate
of the associated uncertainties
it is thus extremely important to develop computing codes which allow for fast
calculations with small numerical uncertainties.

The \hturbo{} program, which is presented in this paper,
aims at providing
fast and numerically precise predictions of the 
Higgs boson production cross sections for phenomenological
applications following the structure of the \dyturbo{} code~\cite{Bozzi:2010xn}
developed for Drell--Yan lepton pair production.
The enhancement in performance over
previous numerical programs is mainly achieved
by introducing one- and
multi-dimensional numerical integrations using
quadrature rules
based on interpolating functions, by software profiling optimization and also by
implementing the multi-threading option.

The \hturbo{} program provides higher-order QCD predictions for the
cross section of
Higgs boson production and decay
at fully differential level in the four momenta of the final states
implementing the \qt{} re\-summation formalism
developed in Refs.~\cite{Catani:2000vq,Bozzi:2005wk,Bozzi:2007pn,Catani:2010pd},
and the \qt{} subtraction method of Ref.~\cite{Catani:2007vq} in a completely
independent way from the original numerical programs
\hqt~\cite{Bozzi:2005wk,deFlorian:2011xf}, \hnnlo~\cite{Catani:2007vq}  and
\hres~\cite{deFlorian:2012mx}.
This novel implementation, other than an improvement in performances
and numerical precision, has the aim of facilitate the inclusion of
N$^3$LO corrections along the lines of Ref.~\cite{Camarda:2021ict} and the
fiducial perturbative power corrections
within the $q_T$ subtraction method
exploiting the recoil procedure of Ref.\,\cite{Catani:2015vma}
as performed in Ref.~\cite{Camarda:2021jsw}.
The present version of the program includes the Higgs boson production
through gluon fusion and its decay in a photon pair,
implementing the resummation of
the log\-a\-rith\-mi\-cal\-ly-enhanced QCD contributions in the small-\qt{}
region of the Higgs boson at leading-logarithmic (LL), next-to-leading-logarithmic (NLL), and
NNLL accuracy, and including the
corresponding finite-order contributions at next-to-leading order
(NLO) and  NNLO both in the small- and large-\qt{} region\,\footnote{Sometimes in the literature
  this is referred respectively as NLL$'$ and  NNLL$'$ accuracy.}.
The fully-differential fixed-order QCD calculation have been implemented up to
next-to-next-to-leading order (NNLO).
The
$H$+jet predictions at $O(\as)$ and $O(\as^2)$ 
have been reimplemented from the
the MCFM
program~\cite{Campbell:2010ff,Boughezal:2016wmq},
as encoded in \hres{} and \hnnlo.
The \hturbo{} program is based on a modular C++ structure
(with few Fortran functions interfaced)
with multi-threading
implemented with OpenMP and through the Cuba
library~\cite{Hahn:2014fua}.
The parameters of the calculation can
be set in a flexible way via input file and/or command line options. 
The \hturbo{} program will be made publicly available.

We briefly summarize the relevant formulae which have been implemented in the \hturbo{} program\,\footnote{The reader
  interested on the details of the \qt{}-resummation and \qt{}-subtraction formalisms
 is referred to the original literature~\cite{Catani:2000vq,Bozzi:2005wk,Bozzi:2007pn,Catani:2010pd,Catani:2007vq}.}.
The fully-differential Higgs boson cross section, completely inclusive over the final-state QCD radiation,
is described by six kinematic variables corresponding to the momenta of the two photons.
Therefore we can expressed the cross-section as a function of the
transverse momentum \qT, the rapidity $y$ and the invariant mass $m$
of the Higgs boson (or photon pair), and three angular variables corresponding to the
polar angle $\theta$ and azimuth $\phi$
of the photon decay in a given boson rest
frame  and to the azimuth $\phi_{H}$ of the Higgs boson in the laboratory frame.
Given the spin-0 nature of the SM Higgs boson
the fully-differential cross section factorizes
in two independent factors for the Higgs boson production and decay subprocesses.
We treat the  Higgs boson within the narrow-width approximation, $\Gamma_H/m_H\to 0$ ($\Gamma_H$ is the Higgs boson total decay width),  and therefore
we have $m=m_H$.
Moreover  in (unpolarised)
hadron collisions the initial-state hadrons, i.e. the incoming beams,
are to very good approximation azimuthally symmetric and therefore the cross section does not depend on the absolute value of $\phi_{H}$.
Therefore in the following we will consider the cross section
averaged over $\phi_{H}$ at
fixed values of the additional kinematical variables of the
final-state system. 

The \qt-resummed cross section 
for Higgs boson production can be written as
\begin{eqnarray}
\label{eq:rescross_1}
\textrm{d}\sigma^{\textrm{H}}_{\textrm{N$^{n}$LL+N$^{n}$LO}}&=&
\textrm{d}\sigma^{\textrm{res}}_{\textrm{N$^{n}$LL}}
-\textrm{d}\sigma^{\textrm{asy}}_{\textrm{N$^{n-1}$LO}}
+\textrm{d}\sigma^{\textrm{f.o.}}_{\textrm{N$^{n-1}$LO}}\, ,
\end{eqnarray}
with $n=1,2,3,\dots$ (in the following we do not explicitly consider the lowest order predictions
at LL accuracy: $\textrm{d}\sigma^{\textrm{H}}_{\textrm{LL}}= \textrm{d}\sigma^{\textrm{res}}_{\textrm{LL}}$).
The term  $\textrm{d}\sigma^{\textrm{res}}$ in Eq.\,(\ref{eq:rescross_1})  is the resummed
component, $\textrm{d}\sigma^{\textrm{asy}}$ is
the asymptotic contribution (that is the fixed-order expansion of
$\textrm{d}\sigma^{\textrm{res}}$), and
$\textrm{d}\sigma^{\textrm{f.o.}}$ is the finite-order
cross section integrated over final-state QCD radiation (which can be obtained from the $H$+jet cross section).
The resummed term $\textrm{d}\sigma^{\textrm{res}}$ dominates at small \qt{} ($\qt\ll m$)
while the finite-order component
$\textrm{d}\sigma^{\textrm{f.o.}}$ correctly describe the large-\qt{}
region ($\qt\sim m$). In order to obtain
an accurate description of the  region of intermediate \qt{} a consistent
matching between resummed and finite component is essential.

The resummation of the logarithmic contributions has been
carried out in the impact-parameter space $b$
(which is the Fourier-conjugate variable to \qt)\,\cite{Parisi:1979se}
in order to fulfill the constraint of transverse-momentum
conservation for multi-parton radiation.
Moreover convolution with PDFs is more
conveniently expressed by considering
double Mellin moments of the corresponding
partonic functions\,\cite{Bozzi:2007pn}.
The resummed and asymptotic terms in Eq.\,(\ref{eq:rescross_1}) can thus
be written as\,\footnote{For the sake of simplicity
we use a symbolic notation where convolution
with PDFs, the sum over different initial-state partonic channels
and the
inverse Mellin and Fourier transformations are understood.}:
\begin{eqnarray}
\label{eq:rescross_2}
\textrm{d}\sigma^{\textrm{res}}_{\textrm{N$^{n}$LL}} &=&
\textrm{d}{\hat \sigma}^{\textrm{H}}_{\textrm{LO}} 
\times \mathcal{H}^{\textrm{H}}_{\textrm{N$^{n}$LO}}
\times \exp\{\mathcal{G}_{\textrm{N$^{n}$LL}}\}\\
\label{eq:resct}
\textrm{d}\sigma^{\textrm{asy}}_{\textrm{N$^{n-1}$LO}} &=&
\textrm{d}{\hat \sigma}^{\textrm{H}}_{\textrm{LO}} 
\times \Sigma^{\textrm{H}}(\qt/Q)_{\textrm{N$^{n-1}$LO}} \,,
\end{eqnarray}
where $Q\sim m$ denotes the so-called resummation scale \cite{Bozzi:2005wk},
an auxiliary scale that is introduced in 
$\textrm{d}\sigma^{\textrm{res}}$ and, consistently, 
in $\textrm{d}\sigma^{\textrm{asy}}$ whose variations can be used to
estimate the uncertainty
from not yet calculated higher-order logarithmic corrections.
 The factor $\textrm{d}{\hat \sigma}^{\textrm{H}}_{\textrm{LO}}$ 
is the Born level cross
section,
the coefficient $\mathcal{H}^{H}$\,\cite{Catani:2011kr} is the
(process dependent) hard-collinear  function and
the term $\exp(\mathcal{G})$ is the gluon Sudakov form
factor\,\cite{Catani:1988vd,deFlorian:2000pr,deFlorian:2001zd}
which resums in an exponential form the large logarithmic corrections
in the impact-parameter space $b$.
The function $\Sigma^{\textrm{H}}(\qT/Q)$ can be obtained from the fixed-order
expansion of the term $\mathcal{H}^{\textrm{H}} \times \exp\{\mathcal{G}\}$,
and it embodies the 
singular behaviour of $\textrm{d}\sigma^{\textrm{f.o.}}$ in the limit $\qT \to 0$.

The Mellin moments of the hard-collinear function
$\mathcal{H}^{H}$ have been computed
with the {\tt FORM}\,\cite{Kuipers:2012rf}
packages {\tt summer}\,\cite{Vermaseren:1998uu}
and {\tt harmpol}\,\cite{Remiddi:1999ew},
using the method of Ref.\,\cite{Albino:2009ci}.
The Mellin space evolution of PDFs and the Mellin moments of the
splitting functions have been
calculated with the package QCD-PEGASUS\,\cite{Vogt:2004ns}.

The \hturbo{} program includes also fixed-order predictions (without the
resummation of log\-a\-rith\-mi\-cal\-ly-enhanced contributions).
 Beyond the LO, the fixed-order cross section is computed through the
 \qt{} subtraction formalism~\cite{Catani:2007vq} and 
 is  expressed as the sum of three terms as follows:
\begin{eqnarray}
  \label{eq:focross_1}
  \textrm{d}\sigma^{\textrm{H}}_{\textrm{N$^{n}$LO}}&=&
  \mathcal{H}^{\textrm{H}}_{\textrm{N$^{n}$LO}}\times\textrm{d}\sigma^{\textrm{H}}_{\textrm{LO}} 
 +\left[\textrm{d}\sigma^{\textrm{H+jet}}_{\textrm{N$^{n-1}$LO}}-\textrm{d}\sigma^{\textrm{CT}}_{\textrm{N$^{n-1}$LO}}\right]\, ,
\end{eqnarray}
 where the term $\textrm{d}\sigma^{\textrm{H+jet}}$ is the $H$+jet cross section,
and the counter-term
$\textrm{d}\sigma^{\textrm{CT}}_{\textrm{N$^{n-1}$LO}}$ is given by
\begin{eqnarray}
  \label{eq:foct} \textrm{d}\sigma^{\textrm{CT}}_{\textrm{N$^{n-1}$LO}} &=&
  \textrm{d}\sigma^{\textrm{H}}_{\textrm{LO}} \times
  \int_{0}^{\infty}  \textrm{d}^2\qt^\prime \, \Sigma^{\textrm{H}}(\qt^\prime/m)_{\textrm{N$^{n-1}$LO}} \,.
\end{eqnarray}
The singular behaviour of $\textrm{d}\sigma^{\textrm{H+jet}}$ in the limit
${\qT} \to 0$, 
known from the $q_T$ resummation
formalism,
is the same of the subtraction counter-term  $\textrm{d}\sigma^{\textrm{CT}}$.
Being the terms $\textrm{d}\sigma^{\textrm{H+jet}}$ and
$\textrm{d}\sigma^{\textrm{CT}}$ in Eq.\,(\ref{eq:focross_1})
separately divergent at $q_T=0$ a technical parameter
$q_T^{\rm cut}>0$ has to be introduced.
For $\qT \geq {\qT}_{\textrm{\scriptsize cut}}$
the sum of the terms in the square bracket of
Eq.~(\ref{eq:focross_1}) is IR finite
(or, more precisely, integrable over $q_T$)
and the ``exact''
value of the cross section can be obtained evaluating 
the square bracket term in
  Eq.~(\ref{eq:focross_1}) in the limit
  ${\qT}_{\textrm{\scriptsize cut}} \to 0$. 
  However for finite value of $q_T^{\rm cut}$ the cross section in
  Eq.\,(\ref{eq:focross_1}) contains perturbative power corrections ambiguities 
  $\mathcal{O}((q_T^{\rm cut}/M)^p)$\,\cite{Cieri:2019tfv,Ebert:2019zkb,Buonocore:2019puv,Oleari:2020wvt}, with $p>0$
  which are particularly severe in the case of fiducial selection cuts
  which yield an acceptance that has a residual linear dependence
on $q_T^{\rm cut}$\,\cite{Ebert:2019zkb,Ebert:2020dfc,Alekhin:2021xcu}.   
A method to remove such linear fiducial  power corrections (FPC) within
the $q_T$ subtraction formalism has been proposed in Refs.\,\cite{Camarda:2021jsw,Buonocore:2021tke}.

An important feature of the resummation
formalism of Ref.\,~\cite{Bozzi:2005wk} is the so called unitarity constraint which leads to 
the following relation:
\begin{eqnarray}
  \label{eq:matching_1} \int_{0}^{\infty} \textrm{d}\qt^2 \, \textrm{d}\sigma^{\textrm{res}}_{\textrm{N$^{n}$LL+N$^{n}$LO}}
  &=& {\cal H}^{\textrm{H}}_{\textrm{N$^{n}$LO}}\times\textrm{d}{\hat \sigma}^{\textrm{H}}_{\textrm{LO}}\,,
\end{eqnarray}
which ensures that fixed-order results are exactly recovered
upon integration over \qt{} of the matched cross section.
A consequence of the unitarity constraint is the reduction of resummation effects in the region of small impact parameter
where it is clear that resummation cannot gives an improvement over the accuracy of the fixed-order calculation. 
The contribution of unjustified resummed contributions  in the large-\qt{} region
can be further reduced or eliminated
by introducing a switching function $w(\qt,m)$ which multiplies the terms $\textrm{d}\sigma^{\textrm{res}}_{\textrm{N$^{n}$LL}}$ and
$\textrm{d}\sigma^{\textrm{asy}}_{\textrm{N$^{n-1}$LO}}$ in Eq.\,(\ref{eq:rescross_1}) above a given \qt{} value suppressing the resummation effects
in the large-\qt{}  region.
However because
such switching violates the unitarity constraint of Eq.~(\ref{eq:matching_1}) 
it has to be included with some care.
Within \hturbo{} the effect of a Gaussian switching
function $w(\qt,m)$ chosen following Ref.~\cite{Catani:2015vma}
can be included.

The perturbative form factor $\exp(\mathcal{G})$
is formally singular for transverse-momenta
of the order of the scale of the Landau pole of the QCD coupling
($b^{-1}\sim \Lambda_{QCD}$) is approached. This is the indication of
the breakdown of perturbation
theory and of the onset of truly non-perturbative (NP) effects. In this region
a model for NP QCD effects,
which has to include a regularization of the Landau singularity is necessary. 
In the \hturbo{} program it  has been explicitly implemented the so-called
{\itshape Minimal Prescription}\,\cite{Catani:1996yz,Laenen:2000de,Kulesza:2002rh} which regularizes the Landau singularity
in resummed calculations  without
introducing  higher-twist power-suppressed contributions of the type
$\mathcal{O}(\Lambda_{QCD}/Q)$.
As alternative it can be chosen the freezing procedure\,\cite{Collins:1981va,Collins:1984kg} known as the `$b_*$ prescription',
which is implemented in \hres{},
consisting in the replacement
\begin{eqnarray}
\label{bstar}
b^2 \to b_*^2 = b^2 \;b_{\rm lim}^2/( b^2 + b_{\rm lim}^2) 
\end{eqnarray}
in the form factor $\exp(\mathcal{G})$. The value of the parameter 
$b_{\rm lim}$ has to be set to be slightly smaller than the Landau singularity
$b^{-1}\sim \Lambda_{QCD}$.
Power-suppressed contributions are expected to dominate 
at very small transverse-momentum ($q_T\sim \Lambda_{QCD}$) and have to be (properly) included
taking into account the delicate interplay with the leading-twist term
in order to correctly describe the experimental data in that region.
We parametrize the NP QCD effects at low \qt{} through a non-perturbative form factor with different
functional forms (the simplest one is a Gaussian smearing factor $\exp{-g_{NP} b^2}$ which depends on the non perturbative
parameter $g_{NP}$).

In the following we show some benchmark numerical results obtained with \hturbo{} compared with
corresponding results from \hres{}
(up to NNLL+NNLO accuracy) and \hnnlo{} (up to NNLO).  In particular we consider the  cross section differential in
the Higgs boson \qt{} in
both the full final state diphoton phase space
and in a given selected fiducial region. We also  compare the time performance of the codes in order to assess
the performance improvement of \hturbo.

\begin{figure*}[]
 \begin{center}
    \subfigure[]{\includegraphics[width=0.495\textwidth]{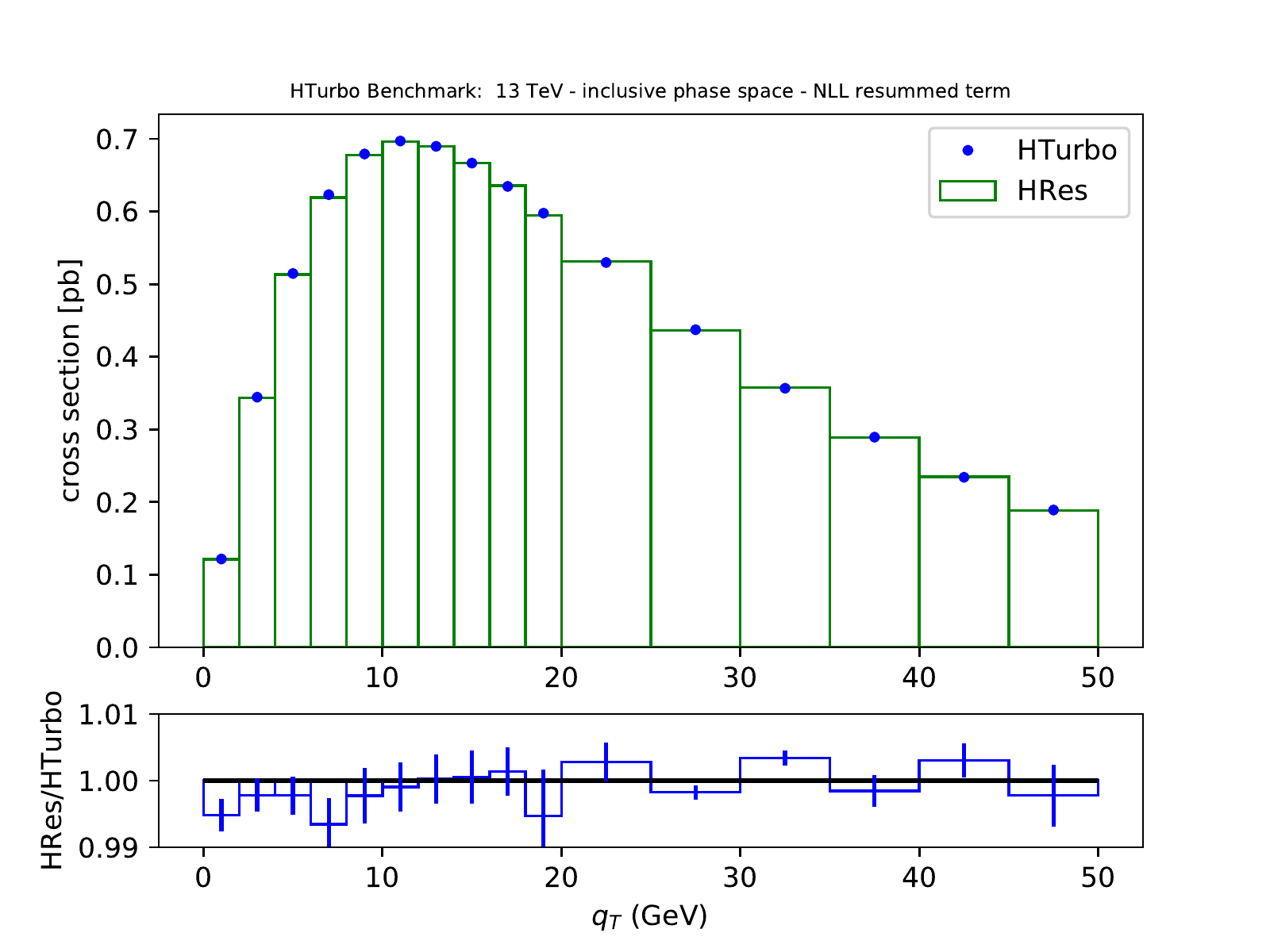}} 
    \subfigure[]{\includegraphics[width=0.495\textwidth]{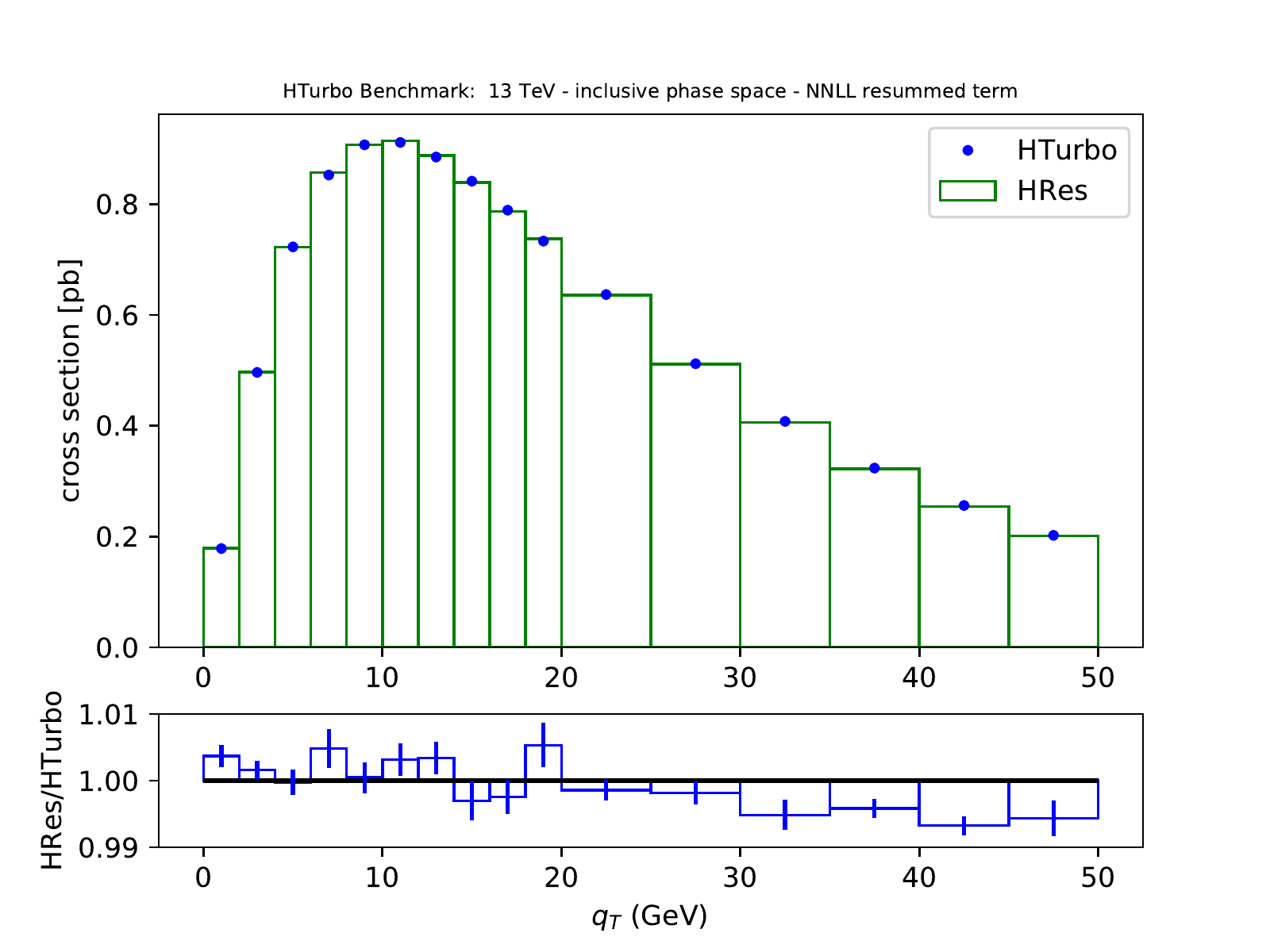}}    \caption{Comparison of full-photon phase space differential cross sections 
      computed with
      \hres{} and \hturbo{} 
      at $\sqrt{s} = 13$~TeV. 
      Resummed component of the transverse momentum distribution at NLL (a) and NNLL (b) accuracy.
      The top panels show absolute cross sections, and the bottom panels show ratios of \hturbo{} to \hres{} results.
 \label{fig:res-nocuts}}
 \end{center}
 \end{figure*}

\begin{figure*}[]
 \begin{center}
    \subfigure[]{\includegraphics[width=0.495\textwidth]{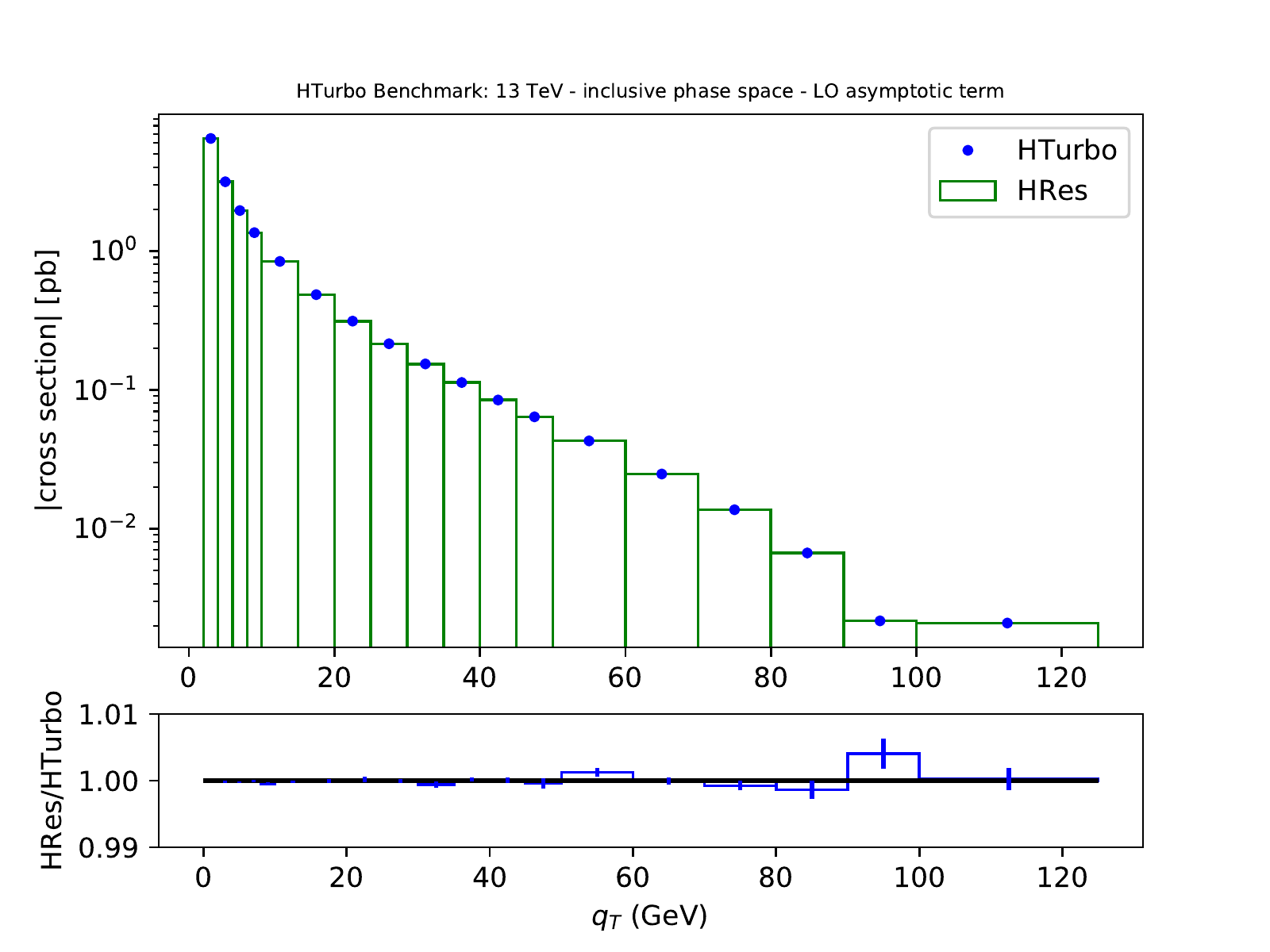}} 
    \subfigure[]{\includegraphics[width=0.495\textwidth]{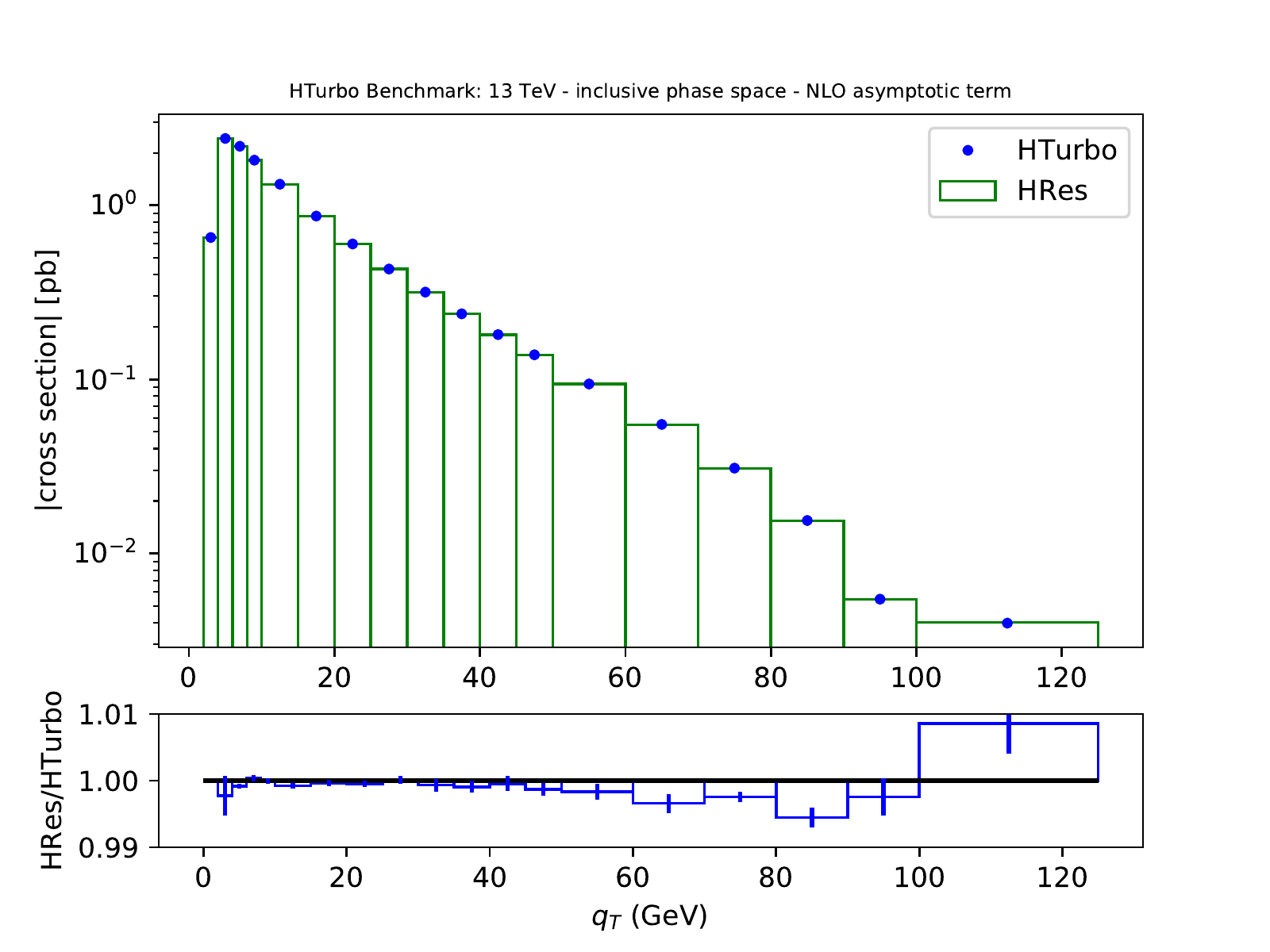}}
   \caption{Comparison of full-photon phase space differential cross sections 
      computed with
      \hres{} and \hturbo{} 
      at $\sqrt{s} = 13$~TeV. 
      Absolute value of the asymptotic component of the transverse momentum distribution at LO (a) and NLO (b) accuracy.
      The top panels show absolute cross sections, and the bottom panels show ratios of \hturbo{} to \hres{} results.
 \label{fig:asy-nocuts}}
 \end{center}
 \end{figure*}

\begin{figure*}[]
 \begin{center}
    \subfigure[]{\includegraphics[width=0.495\textwidth]{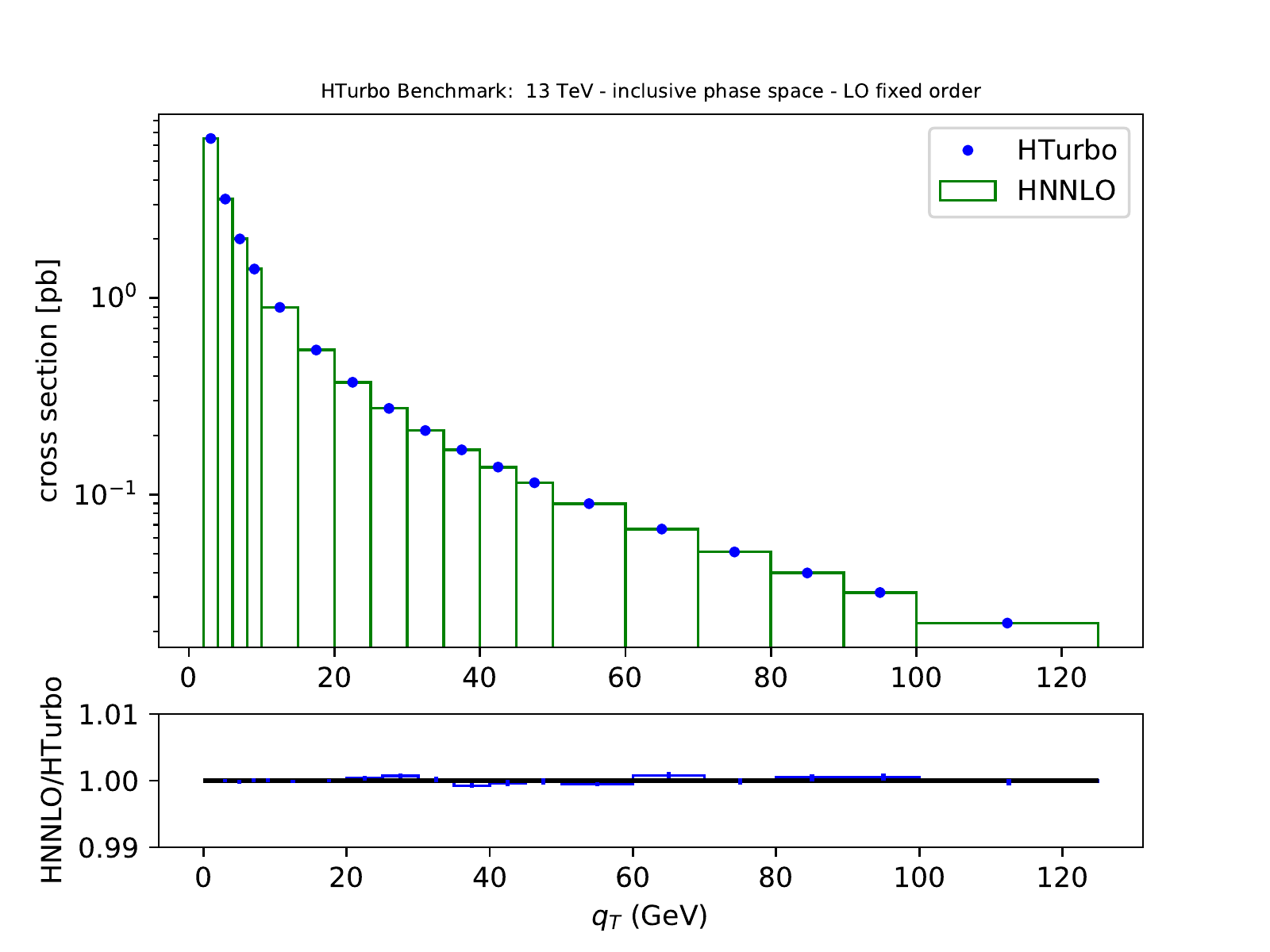}} 
    \subfigure[]{\includegraphics[width=0.495\textwidth]{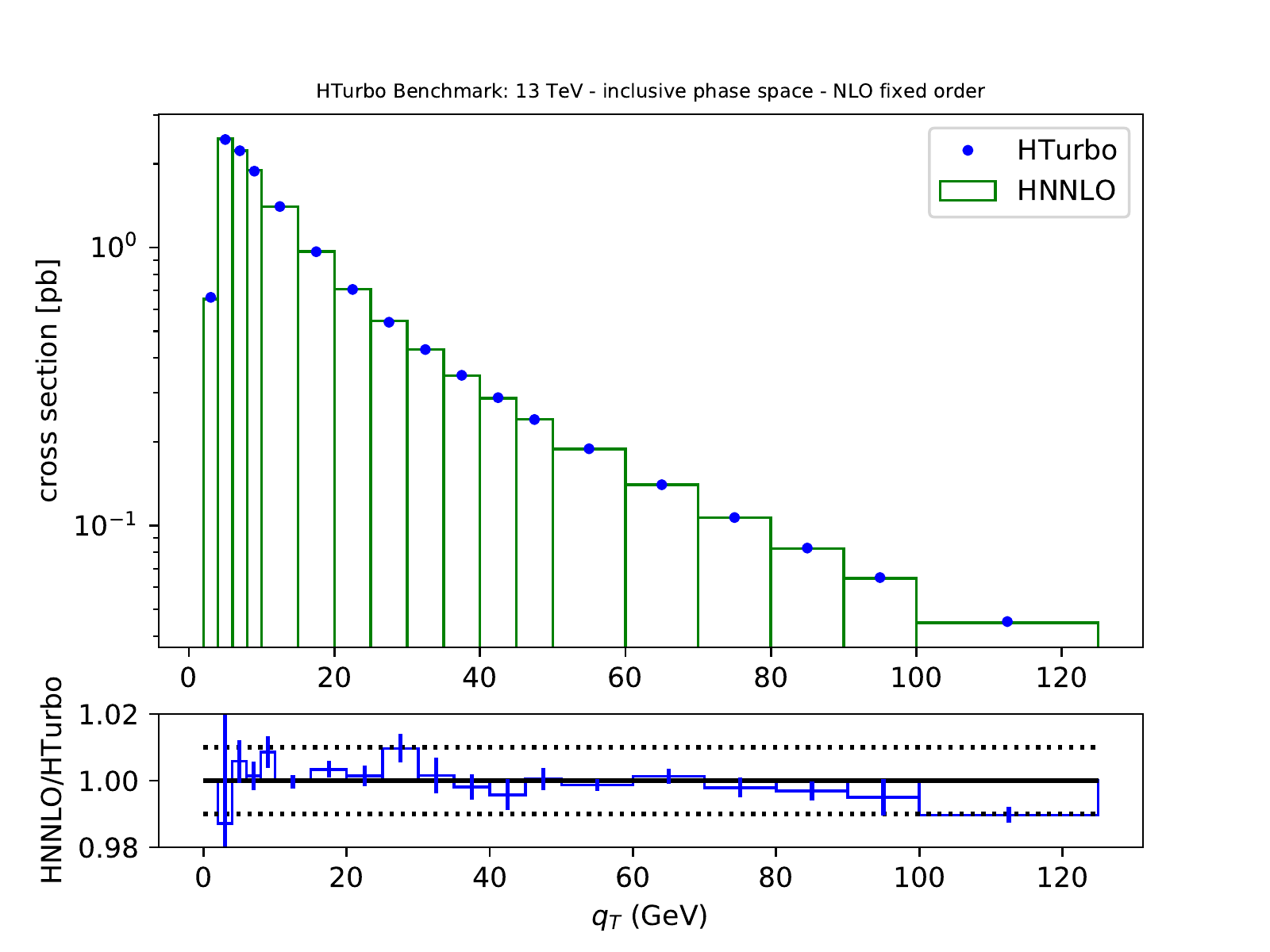}} 
   \caption{Comparison of full-photon phase space differential cross sections 
      computed with
      \hres{} and \hturbo{} 
      at $\sqrt{s} = 13$~TeV. 
      Fixed-order component of the transverse momentum distribution at LO (a) and NLO (b) accuracy.
      The top panels show absolute cross sections, and the bottom panels show ratios of \hturbo{} to \hres{} results.
 \label{fig:fo-nocuts}}
 \end{center}
 \end{figure*}

We consider Higgs boson cross
sections in proton--proton collisions at $\sqrt{s} = 13$~TeV using the
NNPDF3.1 NNLO\,\cite{NNPDF:2017mvq}
set of parton density functions with $\as(m_Z)=0.118$.
The computation is
performed by considering $gg \to H$ production,
through a top-quark loop, in the large-$M_{top}$ approximation.
We use the same settings and input parameters in both the \hturbo{} and \hres{}  codes.
In particular the value of the renormalization ($\mu_R$), factorization ($\mu_F$)
and resummation ($Q$) scales have been chosen to be equal to the Higgs boson mass $m_H$. 
We start to present our benchmark results at inclusive level (i.e.\ integrating over the diphoton final state kinematics).
In Fig.\,\ref{fig:res-nocuts} we consider the resummed part of the \qt{} distribution (see Eq.(\,\ref{eq:rescross_2}))
at NLL accuracy (left panel) and at NNLL accuracy (right panel).
The \hturbo{} results using quadrature integration 
 (blue dots) have been compared with the \hres{} results (green histograms). The lower panels show the ratio between
the results which are in agreement, within the numerical uncertainties of the codes, at better than 1\% level.
In Fig.\,\ref{fig:asy-nocuts} we consider the asymptotic term of the cross section (see Eq.(\,\ref{eq:rescross_2})) at LO (left panel) and
NLO (right panel). The asymptotic term diverges in the $q_T \to 0$ limit and it becomes negative at large $q_T$ (we thus show the absolute value of
the results in logarithmic scale).
The $q_T$ distribution of the asymptotic term has been computed in the range $1\,\mbox{GeV}<q_T<m_H$
and we obtained a sub-percent agreement between \hturbo{} (blue dots) using quadrature integration and \hres{} (green histograms) results.
Finally, in Fig.\,\ref{fig:fo-nocuts}, we show the fixed-order term of the cross section at LO (left panel) and NLO (right panel)
as obtained with \hturbo{} and \hnnlo{}.
Since for this term both \hturbo{} and \hnnlo{} program implements the \vegas{} algorithm for numerical integration
we expect to observe similar results as is confirmed by the sub-percent agreement between \hturbo{} (blue dots) and \hnnlo{} (green histograms) results.
 

\begin{figure*}[]
 \begin{center}
    \subfigure[]{\includegraphics[width=0.495\textwidth]{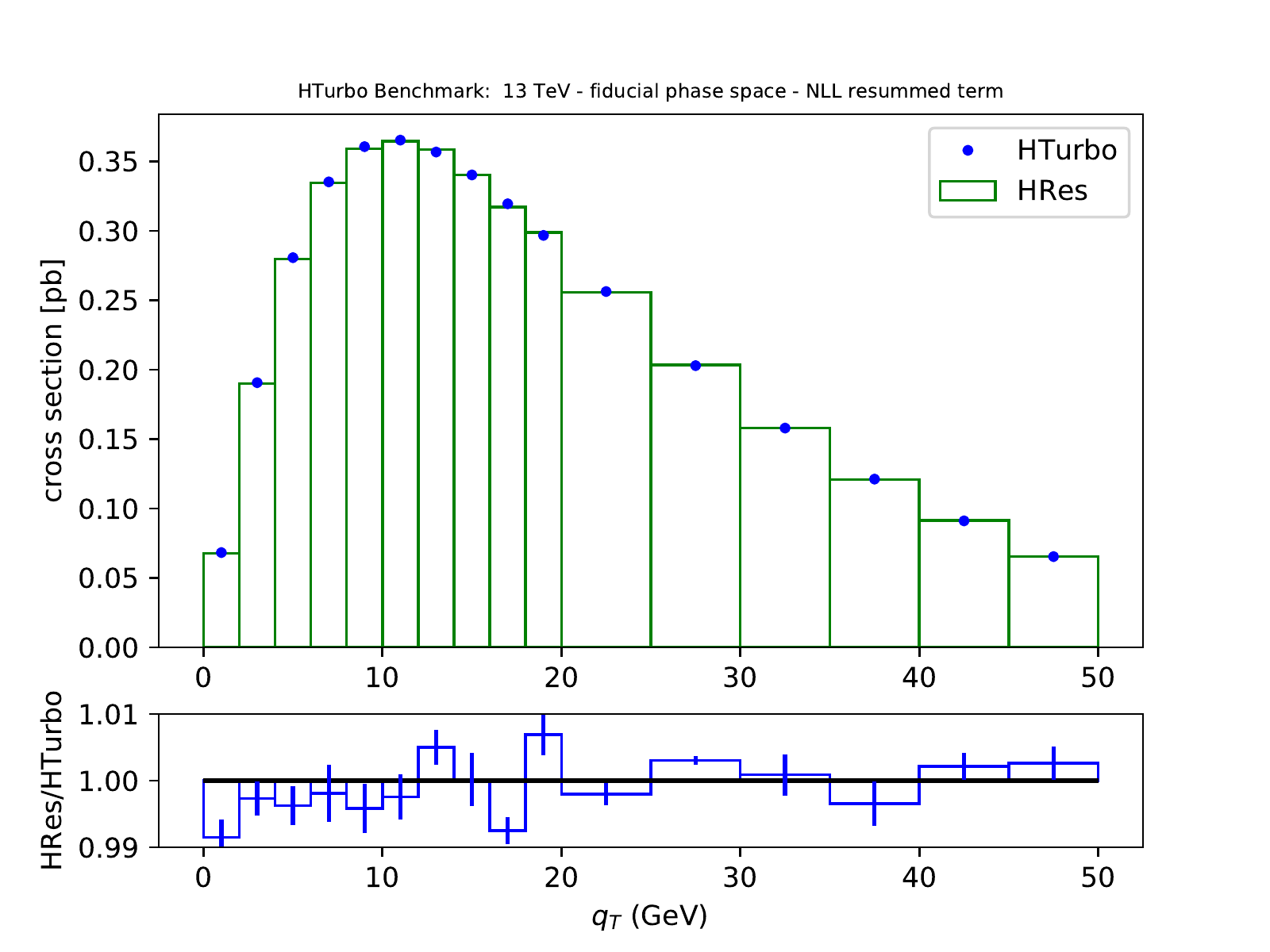}} 
    \subfigure[]{\includegraphics[width=0.495\textwidth]{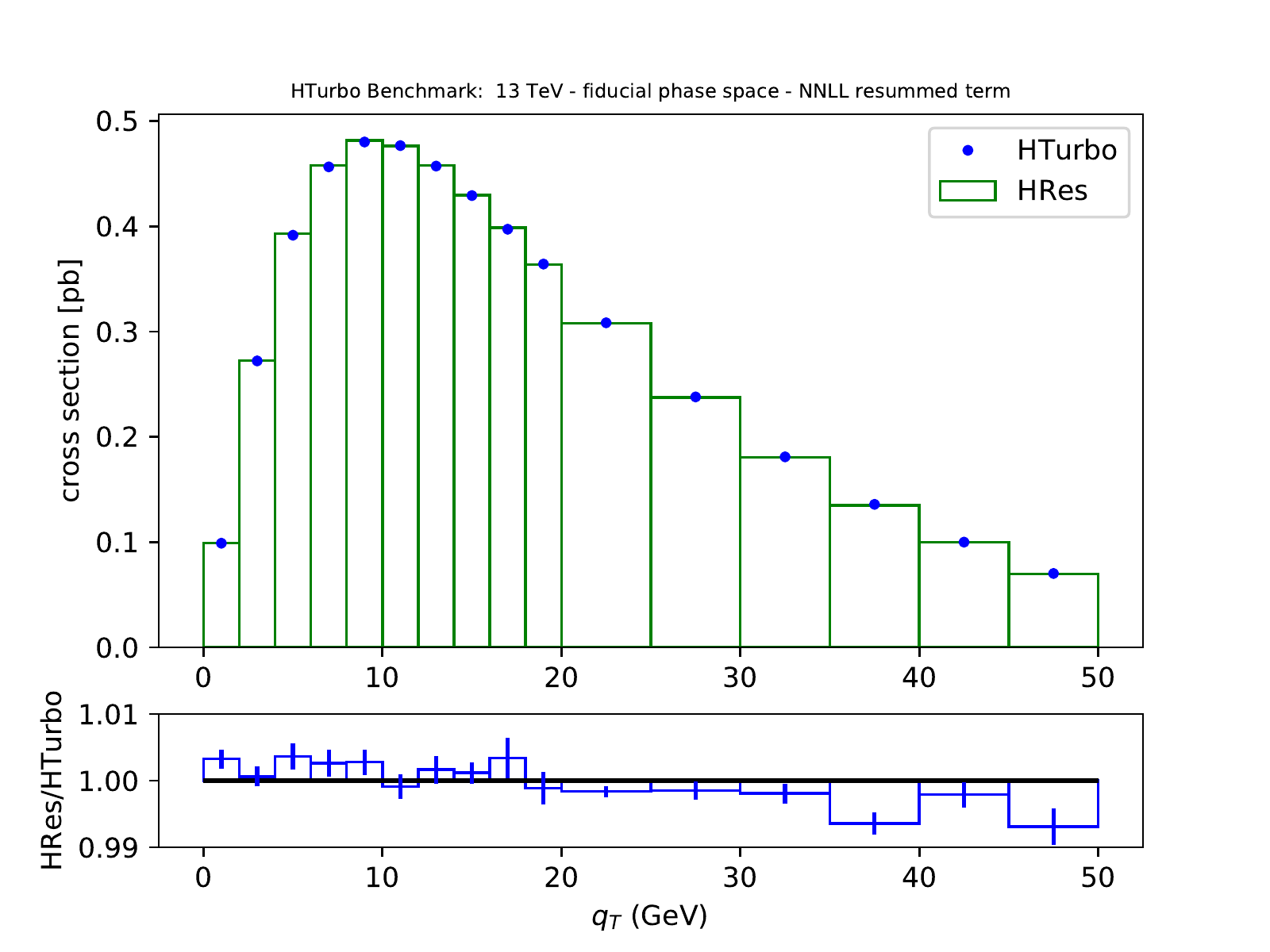}} 
   \caption{Comparison of full-photon phase space differential cross sections 
      computed with
      \hres{} and \hturbo{} 
      at $\sqrt{s} = 13$~TeV. 
      Resummed component of the transverse momentum distribution at NLL (a) and NNLL (b) accuracy.
      The top panels show absolute cross sections, and the bottom panels show ratios of \hturbo{} to \hres{} results.
 \label{fig:res-cuts}}
 \end{center}
 \end{figure*}

\begin{figure*}[]
 \begin{center}
    \subfigure[]{\includegraphics[width=0.495\textwidth]{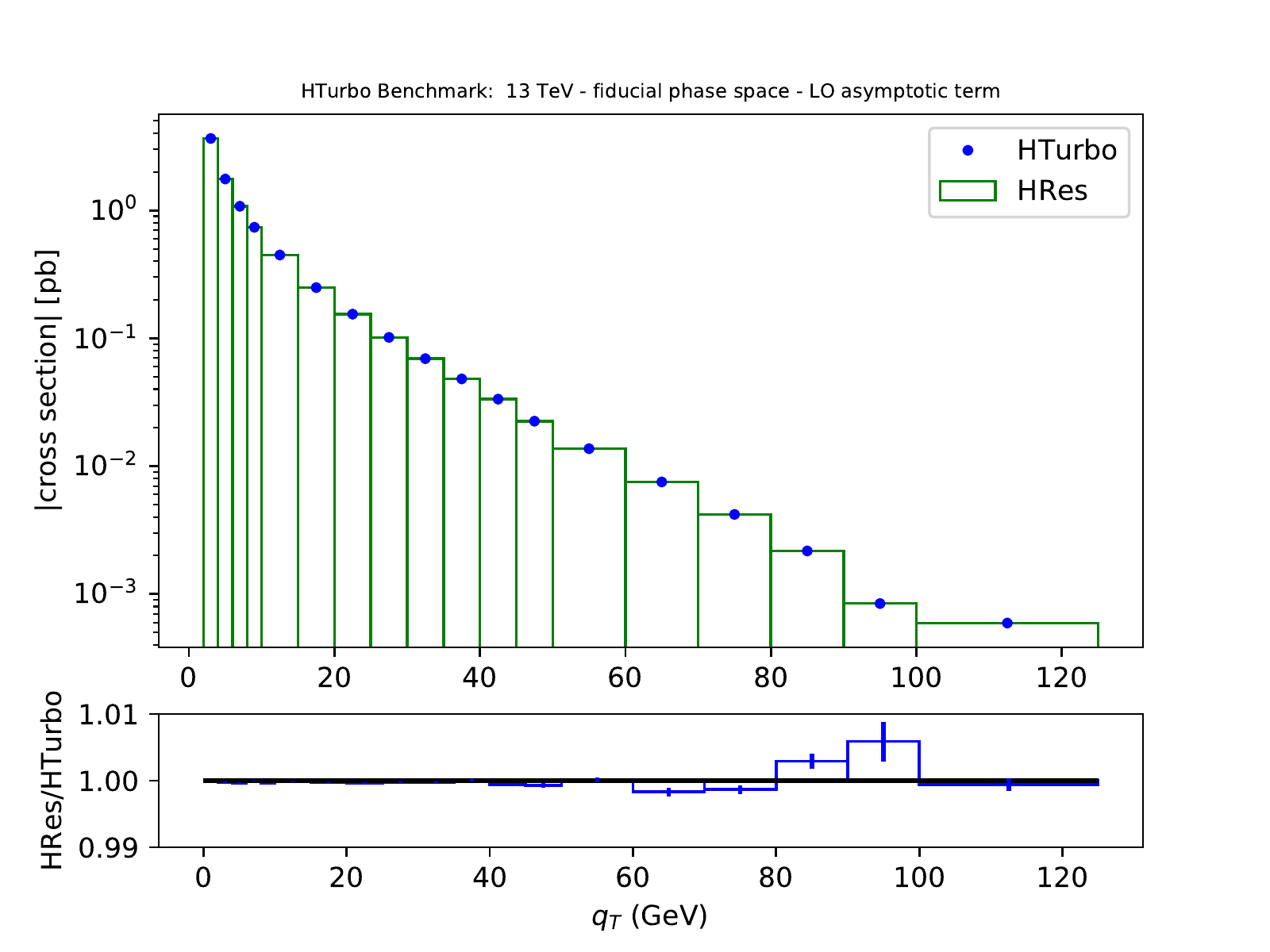}} 
    \subfigure[]{\includegraphics[width=0.495\textwidth]{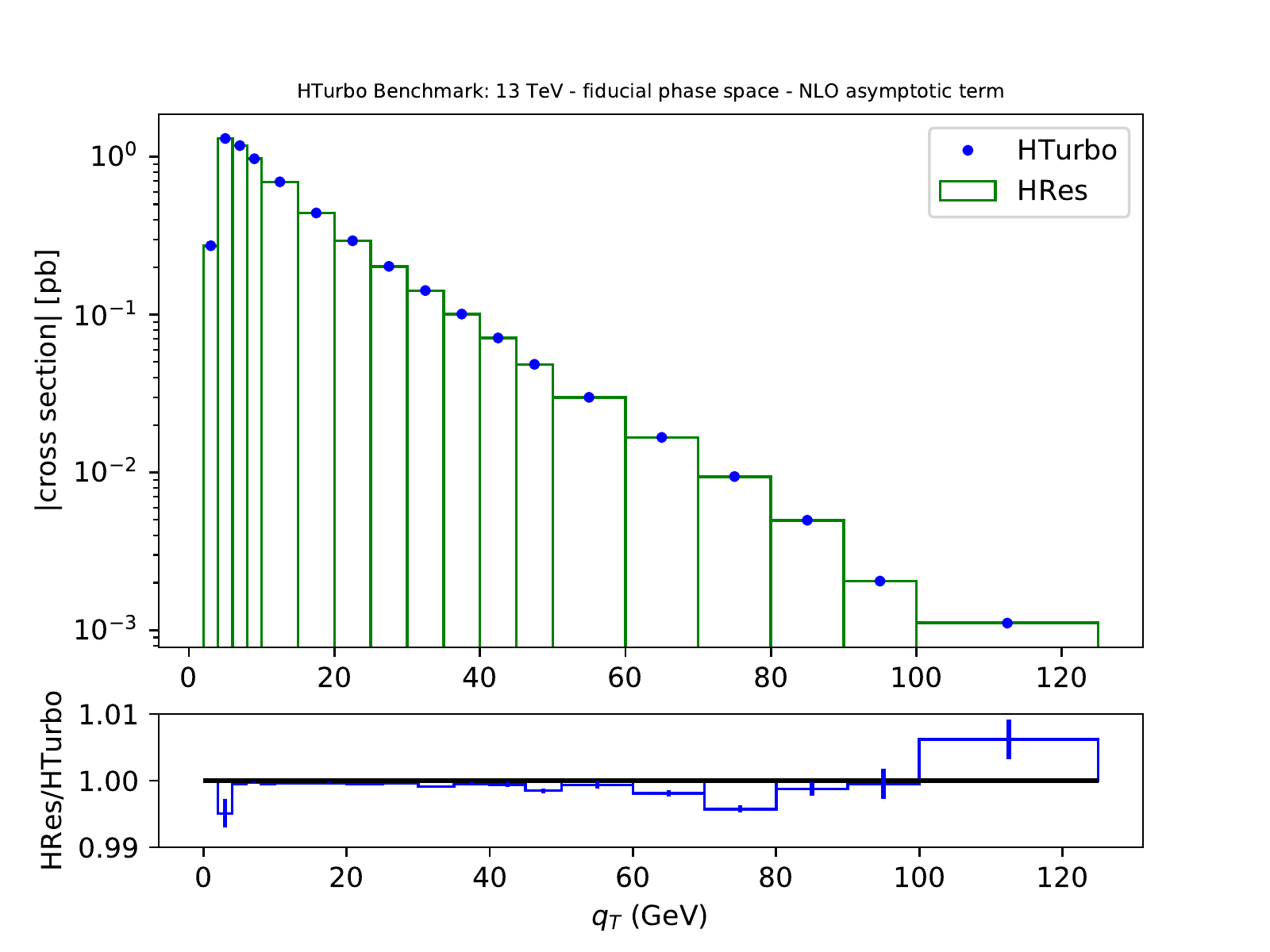}} 
   \caption{Comparison of fiducial differential cross sections 
      computed with
      \hres{} and \hturbo{} 
      at $\sqrt{s} = 13$~TeV. The fiducial phase space is defined in the text.
      Absolute value of the asymptotic component of the transverse momentum distribution at LO (a) and NLO (b) accuracy.
      The top panels show absolute cross sections, and the bottom panels show ratios of \hturbo{} to \hres{} results.
 \label{fig:asy-cuts}}
 \end{center}
 \end{figure*}

\begin{figure*}[]
 \begin{center}
    \subfigure[]{\includegraphics[width=0.495\textwidth]{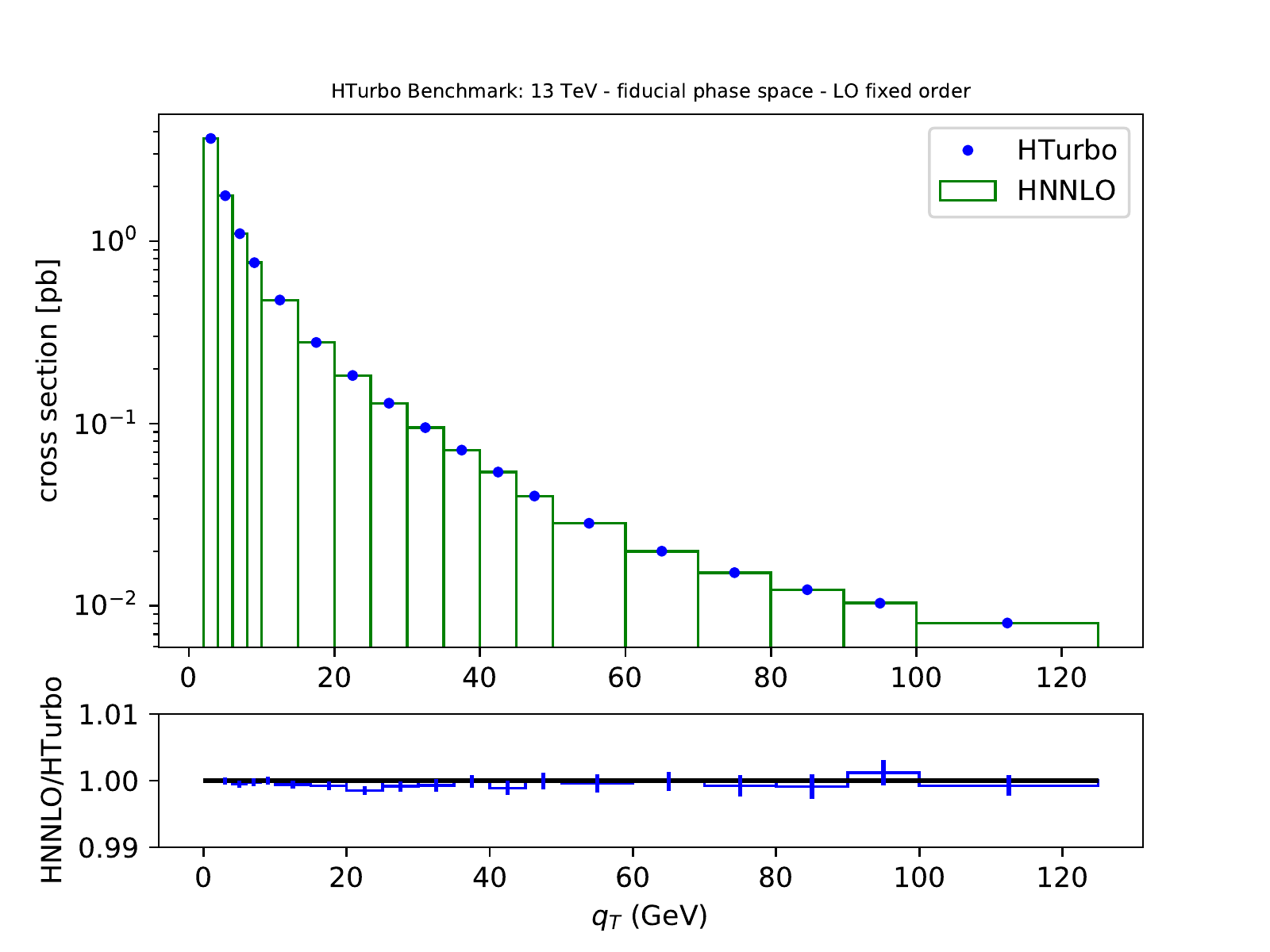}} 
    \subfigure[]{\includegraphics[width=0.495\textwidth]{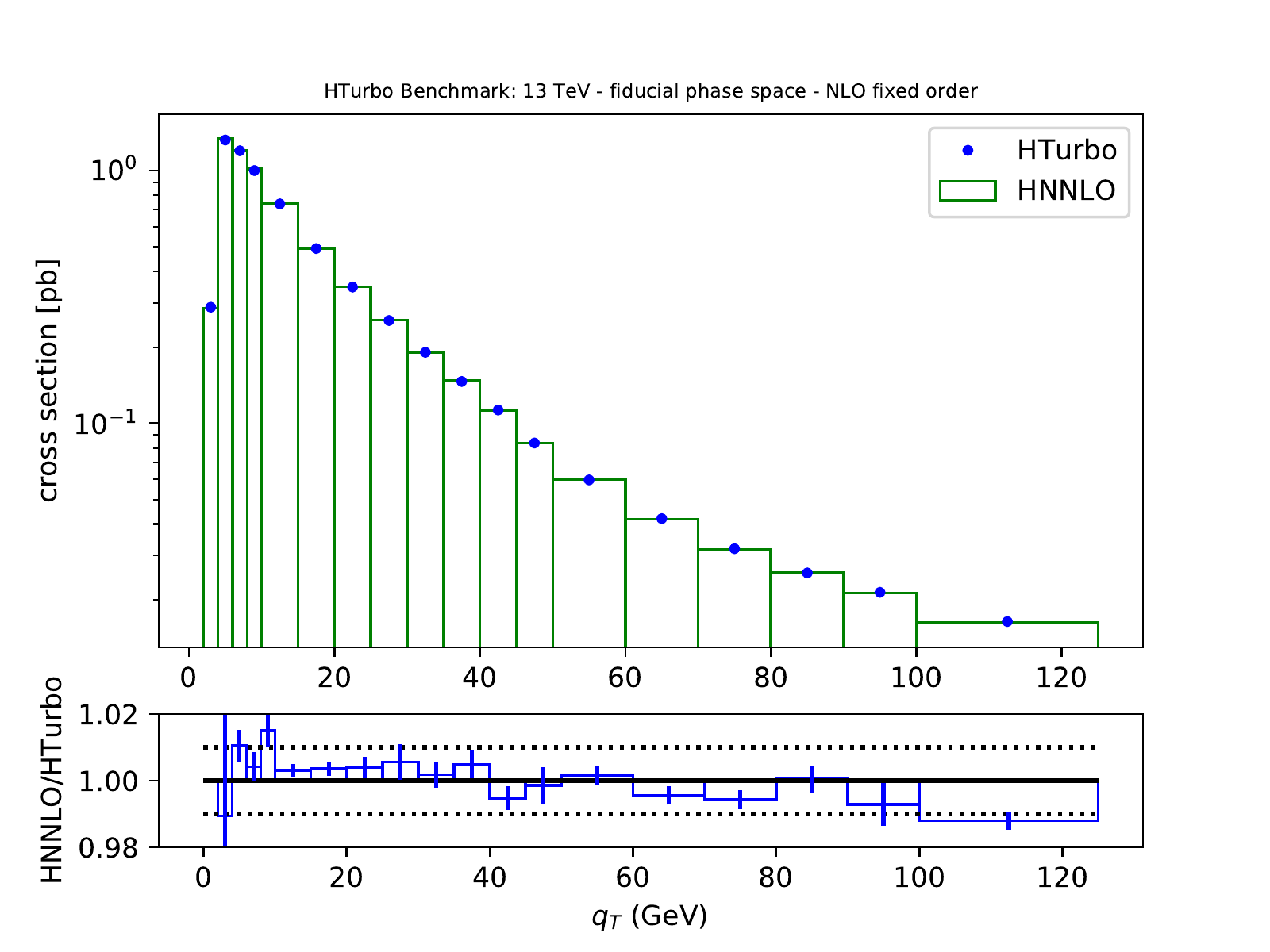}} 
   \caption{Comparison of fiducial differential cross sections 
      computed with
      \hres{} and \hturbo{} 
      at $\sqrt{s} = 13$~TeV. The fiducial phase space is defined in the text.
      Fixed-order component of the transverse momentum distribution at LO (a) and NLO (b) accuracy.
      The top panels show absolute cross sections, and the bottom panels show ratios of \hturbo{} to \hres{} results.
 \label{fig:fo-cuts}}
 \end{center}
 \end{figure*}

We then consider the case of fiducial cross sections.
The fiducial phase space is defined by the photon transverse
momenta $\pT^\gamma>0.35\,m_H$ and the photon pseudorapidities
$|\eta^\gamma|<2.37$. 
In Figs.\,\ref{fig:res-cuts},\ref{fig:asy-cuts},\ref{fig:fo-cuts} we show the comparison between the resummed, asymptotic and fixed-order term as presented
above for the inclusive phase space. Also in this case we observe a sub-percent agreement between
\hturbo{} (blue dots), \hres{} and \hnnlo{} (green histograms) results in the entire range of $q_T$ considered.

We now briefly comment on various tests of time performance
which has been performed on a machine with 3.50 GHz Intel Xeon  CPUs.
The computation time requested to calculate
cross-section predictions for \hturbo{} and \hres{} is
compared and used to assess the performance improvement of \hturbo.
The \hres{} calculation of the resummed term of the inclusive cross-section at NLL accuracy with an uncertainty of 1\% took around 0.5~hour,
while the analogous \hturbo{} calculation (without the multi-threading option)
took around 20~seconds with an uncertainty of 0.001\%, yielding an improvement of around two orders of magnitude
in the time performance together three orders of magnitude in numerical precision.
Similar results were obtained including fiducial cuts or considering the resummed term at NNLL accuracy.
The LO or NLO calculation of the asymptotic term took around 10~minutes with an uncertainty of 0.5-1\% within \hres{} both at inclusive level
and with fiducial cuts, while the analogous \hturbo{} single-thread calculation with a similar accuracy took from 3~seconds (inclusive case) up to 30~seconds,
yielding an improvement of one or two orders of magnitudes in the time performance.
Finally, the fixed-order term of the cross section represents the most time-consuming part of the calculation; the computation
at NLO (LO) with 1\% (0.05\%) accuracy required 30~min (1~min) both within \hnnlo{} and \hturbo{} single-thread.
However we observe that the fixed-order part of the calculation could be computed by using fast interpolation
techniques~\cite{Carli:2010rw,Kluge:2006xs}.

In conclusion, we have presented the \hturbo{} numerical program which provides fast and numerical precise
predictions for Higgs boson production through a new implementation of the
\hqt, \hres{} and \hnnlo{} codes following the improvements of the \dyturbo{} program~\cite{Bozzi:2010xn}
for Drell--Yan lepton pair production.
\hturbo{}  implements the fully-differential fixed-order QCD calculation for Higgs boson production (via gluon fusion)
and decay 
also combined  with the resummation of
the large logarithmic corrections at small transverse momenta.
The present version of the code reaches the next-to-next-to-leading order and next-to-next-to-leading logarithmic
accuracy,
and it includes  the decay of the Higgs boson in two photons.
The enhancement in performance of \hturbo{} with respect to the previous programs (which reaches two orders of magnitude
for the resummed term) 
is achieved by code optimization, by factorizing the cross section into production
and decay variables, and with the usage of numerical integration quadrature rules
based on interpolating functions. The resulting cross-section predictions
are in agreement with the results of the original programs.
The great reduction of computing time for performing cross-sections
calculation opens new possibilities for Higgs boson physics
and facilitate an efficient inclusion of N$^3$LO corrections along the lines of~\cite{Camarda:2021ict}\,.

\paragraph{Acknowledgments.} 
This project has been supported by the European Research Council under the
European Union's Horizon 2020 research and innovation programme
(grant agreement number 740006). LC is supported by the Generalitat
Valenciana (Spain) through the plan GenT program (CIDEGENT/2020/011)
and his work is supported by the Spanish Government (Agencia Estatal de
Investigaci\'on) and ERDF funds from European Commission
(Grant No. PID2020-114473GB-I00 funded by MCIN/AEI/10.13039/501100011033).


\newpage
\begin{thebibliography}{99}

\bibitem{ATLAS:2012yve}
G.~Aad \textit{et al.} [ATLAS],
Phys. Lett. B \textbf{716} (2012), 1-29
doi:10.1016/j.physletb.2012.08.020
[arXiv:1207.7214 [hep-ex]].


\bibitem{CMS:2012qbp}
S.~Chatrchyan \textit{et al.} [CMS],
Phys. Lett. B \textbf{716} (2012), 30-61
doi:10.1016/j.physletb.2012.08.021
[arXiv:1207.7235 [hep-ex]].

  
\bibitem{Anastasiou:2015vya}
C.~Anastasiou, C.~Duhr, F.~Dulat, F.~Herzog and B.~Mistlberger,
Phys. Rev. Lett. \textbf{114} (2015), 212001
doi:10.1103/PhysRevLett.114.212001
[arXiv:1503.06056 [hep-ph]].

\bibitem{Dulat:2018bfe}
F.~Dulat, B.~Mistlberger and A.~Pelloni,
Phys. Rev. D \textbf{99} (2019) no.3, 034004
doi:10.1103/PhysRevD.99.034004
[arXiv:1810.09462 [hep-ph]].

\bibitem{Cieri:2018oms}
L.~Cieri, X.~Chen, T.~Gehrmann, E.~W.~N.~Glover and A.~Huss,
JHEP \textbf{02} (2019), 096
doi:10.1007/JHEP02(2019)096
[arXiv:1807.11501 [hep-ph]].

\bibitem{Billis:2021ecs}
G.~Billis, B.~Dehnadi, M.~A.~Ebert, J.~K.~L.~Michel and F.~J.~Tackmann,
Phys. Rev. Lett. \textbf{127} (2021) no.7, 072001
doi:10.1103/PhysRevLett.127.072001
[arXiv:2102.08039 [hep-ph]].


\bibitem{Chen:2021isd}
X.~Chen, T.~Gehrmann, E.~W.~N.~Glover, A.~Huss, B.~Mistlberger and A.~Pelloni,
Phys. Rev. Lett. \textbf{127} (2021) no.7, 072002
doi:10.1103/PhysRevLett.127.072002
[arXiv:2102.07607 [hep-ph]].



\bibitem{Chen:2014gva}
X.~Chen, T.~Gehrmann, E.~W.~N.~Glover and M.~Jaquier,
Phys. Lett. B \textbf{740} (2015), 147-150
doi:10.1016/j.physletb.2014.11.021
[arXiv:1408.5325 [hep-ph]].

\bibitem{Boughezal:2015dra}
R.~Boughezal, F.~Caola, K.~Melnikov, F.~Petriello and M.~Schulze,
Phys. Rev. Lett. \textbf{115} (2015) no.8, 082003
doi:10.1103/PhysRevLett.115.082003
[arXiv:1504.07922 [hep-ph]].

\bibitem{Boughezal:2015aha}
R.~Boughezal, C.~Focke, W.~Giele, X.~Liu and F.~Petriello,
Phys. Lett. B \textbf{748} (2015), 5-8
doi:10.1016/j.physletb.2015.06.055
[arXiv:1505.03893 [hep-ph]].

\bibitem{Gao:2005iu}
Y.~Gao, C.~S.~Li and J.~J.~Liu,
Phys. Rev. D \textbf{72} (2005), 114020
doi:10.1103/PhysRevD.72.114020
[arXiv:hep-ph/0501229 [hep-ph]].

\bibitem{Cao:2007du}
Q.~H.~Cao and C.~R.~Chen,
Phys. Rev. D \textbf{76} (2007), 073006
doi:10.1103/PhysRevD.76.073006
[arXiv:0704.1344 [hep-ph]].

\bibitem{deFlorian:2012mx}
D.~de Florian, G.~Ferrera, M.~Grazzini and D.~Tommasini,
JHEP \textbf{06} (2012), 132
doi:10.1007/JHEP06(2012)132
[arXiv:1203.6321 [hep-ph]].

\bibitem{Bizon:2018foh}
W.~Bizo\'n, X.~Chen, A.~Gehrmann-De Ridder, T.~Gehrmann, N.~Glover, A.~Huss, P.~F.~Monni, E.~Re, L.~Rottoli and P.~Torrielli,
JHEP \textbf{12} (2018), 132
doi:10.1007/JHEP12(2018)132
[arXiv:1805.05916 [hep-ph]].

\bibitem{Rabemananjara:2020rvw}
T.~R.~Rabemananjara,
JHEP \textbf{12} (2020), 073
doi:10.1007/JHEP12(2020)073
[arXiv:2007.09164 [hep-ph]].

\bibitem{Bozzi:2010xn}
G.~Bozzi, S.~Catani, G.~Ferrera, D.~de Florian and M.~Grazzini,
Phys. Lett. B \textbf{696} (2011), 207-213
doi:10.1016/j.physletb.2010.12.024
[arXiv:1007.2351 [hep-ph]].

\bibitem{Catani:2000vq}
S.~Catani, D.~de Florian and M.~Grazzini,
Nucl. Phys. B \textbf{596} (2001), 299-312
doi:10.1016/S0550-3213(00)00617-9
[arXiv:hep-ph/0008184 [hep-ph]].

\bibitem{Bozzi:2005wk}
G.~Bozzi, S.~Catani, D.~de Florian and M.~Grazzini,
Nucl. Phys. B \textbf{737} (2006), 73-120
doi:10.1016/j.nuclphysb.2005.12.022
[arXiv:hep-ph/0508068 [hep-ph]].

\bibitem{Bozzi:2007pn}
G.~Bozzi, S.~Catani, D.~de Florian and M.~Grazzini,
Nucl. Phys. B \textbf{791} (2008), 1-19
doi:10.1016/j.nuclphysb.2007.09.034
[arXiv:0705.3887 [hep-ph]].

\bibitem{Catani:2010pd}
S.~Catani and M.~Grazzini,
Nucl. Phys. B \textbf{845} (2011), 297-323
doi:10.1016/j.nuclphysb.2010.12.007
[arXiv:1011.3918 [hep-ph]].


\bibitem{Catani:2007vq}
S.~Catani and M.~Grazzini,
Phys. Rev. Lett. \textbf{98} (2007), 222002
doi:10.1103/PhysRevLett.98.222002
[arXiv:hep-ph/0703012 [hep-ph]].


\bibitem{deFlorian:2011xf}
D.~de Florian, G.~Ferrera, M.~Grazzini and D.~Tommasini,
JHEP \textbf{11} (2011), 064
doi:10.1007/JHEP11(2011)064
[arXiv:1109.2109 [hep-ph]].

\bibitem{Camarda:2021ict}
S.~Camarda, L.~Cieri and G.~Ferrera,
Phys. Rev. D \textbf{104} (2021) no.11, L111503
doi:10.1103/PhysRevD.104.L111503
[arXiv:2103.04974 [hep-ph]].

\bibitem{Catani:2015vma}
S.~Catani, D.~de Florian, G.~Ferrera and M.~Grazzini,
JHEP \textbf{12} (2015), 047
doi:10.1007/JHEP12(2015)047
[arXiv:1507.06937 [hep-ph]].

\bibitem{Camarda:2021jsw}
S.~Camarda, L.~Cieri and G.~Ferrera,
[arXiv:2111.14509 [hep-ph]].

\bibitem{Campbell:2010ff}
J.~M.~Campbell and R.~K.~Ellis,
Nucl. Phys. B Proc. Suppl. \textbf{205-206} (2010), 10-15
doi:10.1016/j.nuclphysbps.2010.08.011
[arXiv:1007.3492 [hep-ph]].

\bibitem{Boughezal:2016wmq}
R.~Boughezal, J.~M.~Campbell, R.~K.~Ellis, C.~Focke, W.~Giele, X.~Liu, F.~Petriello and C.~Williams,
Eur. Phys. J. C \textbf{77} (2017) no.1, 7
doi:10.1140/epjc/s10052-016-4558-y
[arXiv:1605.08011 [hep-ph]].

\bibitem{Hahn:2014fua}
T.~Hahn,
J. Phys. Conf. Ser. \textbf{608} (2015) no.1, 012066
doi:10.1088/1742-6596/608/1/012066
[arXiv:1408.6373 [physics.comp-ph]].

\bibitem{Parisi:1979se}
G.~Parisi and R.~Petronzio,
Nucl. Phys. B \textbf{154} (1979), 427-440
doi:10.1016/0550-3213(79)90040-3



\bibitem{Catani:2011kr}
S.~Catani and M.~Grazzini,
Eur. Phys. J. C \textbf{72} (2012), 2013
[erratum: Eur. Phys. J. C \textbf{72} (2012), 2132]
doi:10.1140/epjc/s10052-012-2013-2
[arXiv:1106.4652 [hep-ph]].

\bibitem{Catani:1988vd}
S.~Catani, E.~D'Emilio and L.~Trentadue,
Phys. Lett. B \textbf{211} (1988), 335-342
doi:10.1016/0370-2693(88)90912-4

\bibitem{deFlorian:2000pr}
D.~de Florian and M.~Grazzini,
Phys. Rev. Lett. \textbf{85} (2000), 4678-4681
doi:10.1103/PhysRevLett.85.4678
[arXiv:hep-ph/0008152 [hep-ph]].

\bibitem{deFlorian:2001zd}
D.~de Florian and M.~Grazzini,
Nucl. Phys. B \textbf{616} (2001), 247-285
doi:10.1016/S0550-3213(01)00460-6
[arXiv:hep-ph/0108273 [hep-ph]].

\bibitem{Kuipers:2012rf}
J.~Kuipers, T.~Ueda, J.~A.~M.~Vermaseren and J.~Vollinga,
Comput. Phys. Commun. \textbf{184} (2013), 1453-1467
doi:10.1016/j.cpc.2012.12.028
[arXiv:1203.6543 [cs.SC]].

\bibitem{Vermaseren:1998uu}
J.~A.~M.~Vermaseren,
Int. J. Mod. Phys. A \textbf{14} (1999), 2037-2076
doi:10.1142/S0217751X99001032
[arXiv:hep-ph/9806280 [hep-ph]].

\bibitem{Remiddi:1999ew}
E.~Remiddi and J.~A.~M.~Vermaseren,
Int. J. Mod. Phys. A \textbf{15} (2000), 725-754
doi:10.1142/S0217751X00000367
[arXiv:hep-ph/9905237 [hep-ph]].

\bibitem{Albino:2009ci}
S.~Albino,
Phys. Lett. B \textbf{674} (2009), 41-48
doi:10.1016/j.physletb.2009.02.053
[arXiv:0902.2148 [hep-ph]].

\bibitem{Vogt:2004ns}
A.~Vogt,
Comput. Phys. Commun. \textbf{170} (2005), 65-92
doi:10.1016/j.cpc.2005.03.103
[arXiv:hep-ph/0408244 [hep-ph]].



\bibitem{Cieri:2019tfv}
L.~Cieri, C.~Oleari and M.~Rocco,
Eur. Phys. J. C \textbf{79} (2019) no.10, 852
doi:10.1140/epjc/s10052-019-7361-8
[arXiv:1906.09044 [hep-ph]].


\bibitem{Ebert:2019zkb}
M.~A.~Ebert and F.~J.~Tackmann,
JHEP \textbf{03} (2020), 158
doi:10.1007/JHEP03(2020)158
[arXiv:1911.08486 [hep-ph]].

\bibitem{Buonocore:2019puv}
L.~Buonocore, M.~Grazzini and F.~Tramontano,
Eur. Phys. J. C \textbf{80} (2020) no.3, 254
doi:10.1140/epjc/s10052-020-7815-z
[arXiv:1911.10166 [hep-ph]].


\bibitem{Oleari:2020wvt}
C.~Oleari and M.~Rocco,
Eur. Phys. J. C \textbf{81} (2021) no.2, 183
doi:10.1140/epjc/s10052-021-08878-3
[arXiv:2012.10538 [hep-ph]].


\bibitem{Ebert:2020dfc}
M.~A.~Ebert, J.~K.~L.~Michel, I.~W.~Stewart and F.~J.~Tackmann,
JHEP \textbf{04} (2021), 102
doi:10.1007/JHEP04(2021)102
[arXiv:2006.11382 [hep-ph]].

\bibitem{Alekhin:2021xcu}
S.~Alekhin, A.~Kardos, S.~Moch and Z.~Tr\'ocs\'anyi,
Eur. Phys. J. C \textbf{81} (2021) no.7, 573
doi:10.1140/epjc/s10052-021-09361-9
[arXiv:2104.02400 [hep-ph]].

\bibitem{Buonocore:2021tke}
L.~Buonocore, S.~Kallweit, L.~Rottoli and M.~Wiesemann,
[arXiv:2111.13661 [hep-ph]].

\bibitem{Catani:1996yz}
S.~Catani, M.~L.~Mangano, P.~Nason and L.~Trentadue,
Nucl. Phys. B \textbf{478} (1996), 273-310
doi:10.1016/0550-3213(96)00399-9
[arXiv:hep-ph/9604351 [hep-ph]].

\bibitem{Laenen:2000de}
E.~Laenen, G.~F.~Sterman and W.~Vogelsang,
Phys. Rev. Lett. \textbf{84} (2000), 4296-4299
doi:10.1103/PhysRevLett.84.4296
[arXiv:hep-ph/0002078 [hep-ph]].

\bibitem{Kulesza:2002rh}
A.~Kulesza, G.~F.~Sterman and W.~Vogelsang,
Phys. Rev. D \textbf{66} (2002), 014011
doi:10.1103/PhysRevD.66.014011
[arXiv:hep-ph/0202251 [hep-ph]].

\bibitem{Collins:1981va}
J.~C.~Collins and D.~E.~Soper,
Nucl. Phys. B \textbf{197} (1982), 446-476
doi:10.1016/0550-3213(82)90453-9



\bibitem{Collins:1984kg}
J.~C.~Collins, D.~E.~Soper and G.~F.~Sterman,
Nucl. Phys. B \textbf{250} (1985), 199-224
doi:10.1016/0550-3213(85)90479-1

\bibitem{NNPDF:2017mvq}
R.~D.~Ball \textit{et al.} [NNPDF],
Eur. Phys. J. C \textbf{77} (2017) no.10, 663
doi:10.1140/epjc/s10052-017-5199-5
[arXiv:1706.00428 [hep-ph]].


\bibitem{Carli:2010rw}
T.~Carli, D.~Clements, A.~Cooper-Sarkar, C.~Gwenlan, G.~P.~Salam, F.~Siegert, P.~Starovoitov and M.~Sutton,
Eur. Phys. J. C \textbf{66} (2010), 503-524
doi:10.1140/epjc/s10052-010-1255-0
[arXiv:0911.2985 [hep-ph]].

\bibitem{Kluge:2006xs}
T.~Kluge, K.~Rabbertz and M.~Wobisch,
doi:10.1142/9789812706706\_0110
[arXiv:hep-ph/0609285 [hep-ph]].

\end{thebibliography}
\end{document}